\documentclass[11pt]{article}
\usepackage{geometry}                % See geometry.pdf to learn the layout options. There are lots.
\usepackage{graphicx}
\usepackage{amssymb}
\usepackage{epstopdf}
\usepackage{amsmath}
\usepackage{fixmath}
\usepackage{MnSymbol}
\usepackage{amsfonts}
\usepackage{bm}
\title{Humblonium: Classical Atoms and the Earnshaw Plasma}
\author{Clifford Chafin}
\begin{document}

\maketitle

\begin{abstract}
It is shown that electrostatic and diamagnetic forces can combine to give long lasting metastable bound dimers of macro and mesoscopically sized objects for a physically attainable material regime. 
This can be a large enough effect to support itself in a trap against Earth's gravity and they can stable at very high temperatures.  For a more restricted material parameter set, we investigate the possibility of stable many particle collections that lose their identity as bound pairs and create a kind of plasma.  
These would constitute a kind of transitional state between fluids and granular materials but, unlike usual approaches, the fluid is a gas rather than a liquid.  
%We investigate several configurations and their possible applications to the study of classical complex systems.  Particular interest is paid to the case of nanoparticles which may exhibit some quantum mechanical delocalization over time and let us study transitional effects of quantum and classical behavior.  Methods from the study of ultracold atomic gases are discussed.
\end{abstract}

\section{Introduction}

The stability of matter was a puzzle of the nineteenth century.  It was already believed that collections of positive and negative charges were combined to give net neutral matter in some integer ratios in the form of small constituents.  It was also assumed that Newton's laws reigned at all scales.  Indeed it was Newton's triumph to show that the laws of the cosmos were the same as on Earth so that extension to smaller scales was believed.  A problem arose when Earnshaw showed that there was no stable way to arrange charges or magnets against collapse \cite{Earnshaw}.  The ultimate resolution of this had to await the 1920's and quantum mechanics when the oscillations of the electron wavefunction generated enough quantum pressure to provide stability to planets, stones and atoms.  

There are dynamic and active approaches to magnetic levitation in the form of the Levitron, magnetic bearings and the like.  To achieve static levitation there is a workaround in the form of diamagnetism.  The very tiny diamagnetic forces most notably, in graphite and bismuth, can provide enough force for a permanent magnet to levitate small objects in Earth's gravity.  Superconductors expel all magnetic field so, while they are not diamagnetic per se, they can generate strong levitating forces.  In the case of Type II superconductors, flux pinning can cause this levitation to be stable.  Once again, stable support of bodies requires a quantum effect.  However, unlike the case of bulk matter, the quantum fields need not fill the void between the objects and the stabilization is mediated only by the electromagnetic fields themselves.  It is known that no analog to diamagnetism exists for dielectric response \cite{LL} so magnetic methods are essential.  

It would be quite interesting if we could combine electric and magnetic fields to give bound states of positive and negative pairs that exist as static states of macroscopically sized bodies.  These would be separated enough to be noninteracting in a thermal sense (except by radiation) so that the thermal motions of the bodies would not be damped by the internal atomic motions of the bodies as happens with granular liquids.  Additionally, there would be no delocalization of the bodies necessary to create this support.  As such, these entities would be the closest approximation to classical atoms.  The practical uses of these would depend on what ranges of forces could be manifested, the lifetime of them as (presumably) metastable entities, and if they could exist in large scale collections to define a fluid or granularly obstructed material with no rapid damping in collisions and a possibly tunable angle of repose.  

The purpose of this article is to demonstrate that such pairs are possible and that there is some reason to believe that reasonably stable large collections are as well.  % The thermal velocity of these pairs at most attainable temperatures will be such that they will generally be unbound, so a plasma, but have strong enough diamagnetic repulsion that they will not have hard surface collisions so are still only interacting by the long range fields between them. 
In the case of pairs we dub them ``Humblonium'' in homage to their lowly status as classical bound states with no important quantum character in the regions between the constituents.  
We will demonstrate that micron sized magnetite and superconducting spheres with sustainable bulk charge can give bound pairs strong enough to support themselves against gravity.  Some estimates for collections are considered.  Unlike true atoms and molecules, the distinction of bound pairs or other subcollections are immediately lost leading to an ``Earnshaw plasma'' of mesoscopic particles of uncertain lifetime.  

%The net binding between the collections will sometimes be possible while the pairwise collision is not yielding a unique phase that has attributes of plasma and a surface tension bound liquid with zero vapor pressure.  

\section{Bound Pairs}
\begin{figure}[h!]
   \centering
   \includegraphics[width=3in,trim=0mm 160mm 0mm 50mm,clip]{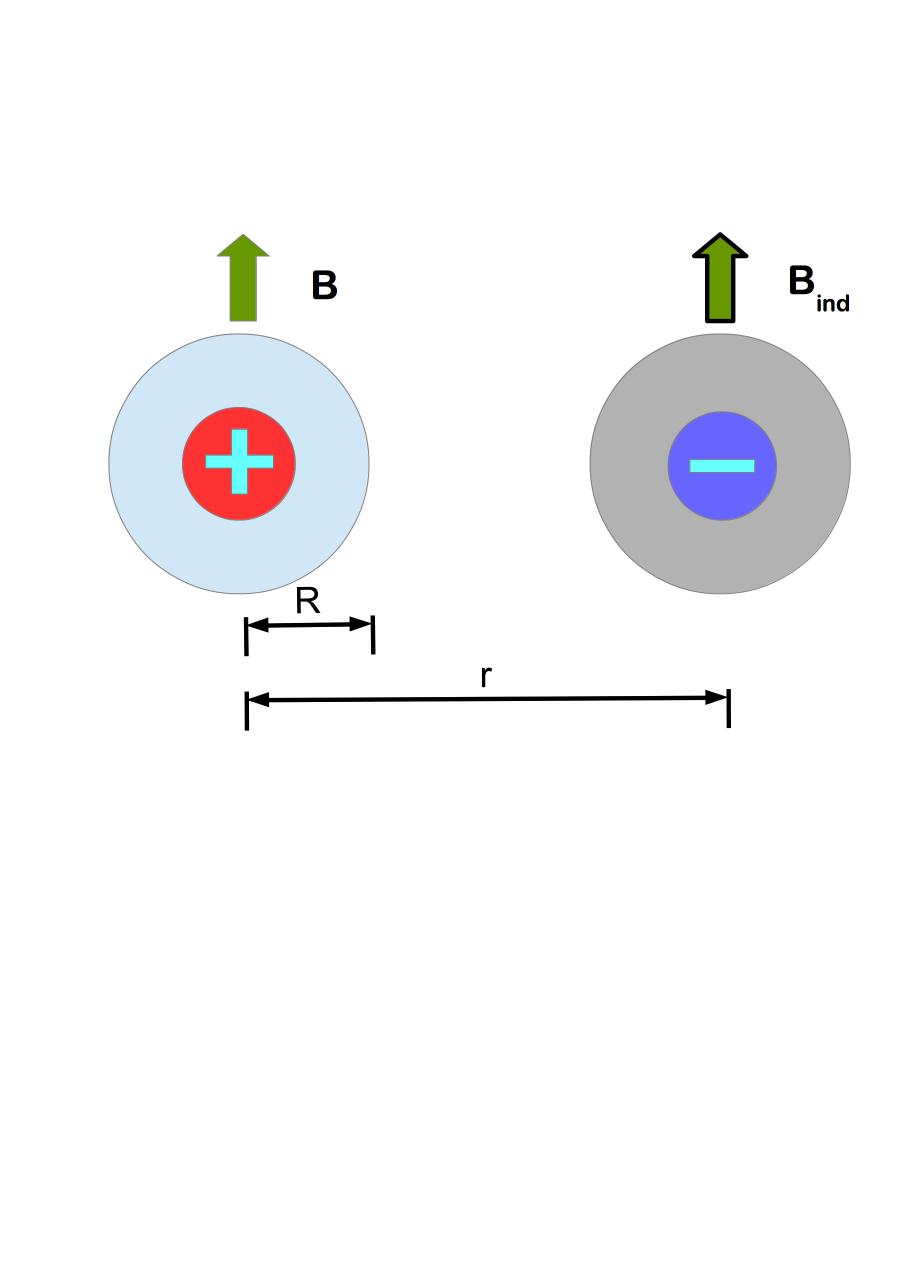} 
   \caption{Electrostatically bound magnet-superconductor pair.}
   \label{balls}
\end{figure}

Consider the case of two equal sized spheres of radius $R$ as in fig.\!~\ref{balls}.  One of them is a strong ferromagnet with surface field $B_{S}$ and opposite charges $\pm q=\pm n e$.  The second sphere is a Type I superconductor which excludes all fields.  The charges are assumed to be localized in domains so they cannot move freely within the balls in response to external fields.  

For these to give a bound state we need the magnetic repulsion to provide enough close repulsion to overcome the electrostatic attraction and the induced ion-induced dipole attraction and the work function to remove the electrons be more than the potential energy to bring the electrons in from infinity.  For these to be of interest we also desire the repulsion to be stronger than the gravitational forces that would pull them downwards in a trap.  Finally, we would like the bound states to be deep enough that they can exist at interesting temperatures.  If the bound states are several times the radii in separation, $r\gg R$, then we can approximate the magnetic repulsion by volume of excluded magnetic energy from the presence of the superconducting sphere.  This separation also facilitates a similar approximation of the ion-induced dipole energy.  

The conditions we seek to satisfy can then be expressed in the following equations
\begin{align}
U_{H}+U_{E}\approx \frac{1}{2}\left(\frac{4\pi R^{3}}{3}\frac{1}{\mu_{0}}\left(\frac{B_{S}R^{3}}{r^{3}}\right)^{2}\right)-\frac{1}{4\pi \epsilon_{0}}\frac{n^{2} e^{2}}{r}\text{~~~~~~stationary}
\end{align}
\begin{align}
\frac{1}{4\pi \epsilon_{0}}\frac{n e^{2}}{R}<U_{\text{Work}}
\end{align}
\begin{align}
0\approx F_{E}+F_{g}=-\frac{1}{4\pi \epsilon_{0}}\frac{n^{2} e^{2}}{r^{2}}+\rho\frac{4\pi R^{3}}{3} g
\end{align}
\begin{align}
U_{H}+U_{E}\ll k_{B}T_{\text{room}}
\end{align}
\begin{align}
U_{\text{induce dipole}}\ll U_{H}
\end{align}

Setting the binding energy of the last electron equal to the work function, extremizing the binding energy and choosing the electric repulsion to repulsion to be equal to the force of gravity we find:
\begin{align}
R&=\frac{1}{2}\left( \frac{3^{5}\epsilon_{0}^{7}\mu_{0}^{2}}{g^{5}\rho^{5}e^{14}U^{2}B_{S}^{4}} \right)^{1/19}\\
r&=\left( \frac{ 3^{7}\epsilon_{0}^{6}U^{12}B_{S}^{2}}{\mu_{0}e^{12}g^{7}\rho^{7}} \right)^{1/19}\\
n&=4\pi \left( \frac{3^{5}\epsilon_{0}^{26}\mu_{0}^{2}U^{33}}{g^{5}\rho^{5}e^{52}B_{S}^{4}} \right)^{1/19}\\
\end{align}

Consider the case of a magnetite sphere with $\rho =5175$kg/m$^{3}$, $\epsilon_{m}\approx 50$, and a surface field of $B_{S}\approx1$T.  Approximating the work function $U =1$eV and using Earth's gravity, $g=9.8$m/s, we obtain  $R=1.56\mu$m, $r=18.3\mu$m, and $n=1086$.  The ratio $r/R=11.7$ so our approximation of $r\gg R$ is valid.  The net binding energy is $U_{net}=77.4$eV so the temperature for  dissociation is $T_{b}=8.97\times10^{5}$K.  Taking 
\begin{align}
U_{\text{induce dipole}}\approx\epsilon_{m}\frac{\epsilon_{0}}{2}\frac{4\pi R^{3}}{3}E(r)^{2}
\end{align}
we see $U_{\text{induce dipole}}=0.29$eV whereas $U_{H}=$15.5eV so that it can be safely ignored.  The thermal velocity is conveniently small, $\sim 1$cm/s and the separation of of the motion and rotation of the spheres from the internal motions means it can presumably persist for a long time despite being higher than any existing temperature for condensed matter.  

%We can consider charged pairs where one of the particles is a lone electron and the other is a superconductor.  The low mass of the electron suggests it can delocalize over the ground state of the potential sphere on a short time scale.  Here we assume that the bound state is close to the superconducting balls o that 
%\begin{figure}[h!]
%   \centering
%   \includegraphics[width=3in,trim=0mm 160mm 0mm 50mm,clip]{pair.jpg} 
%   \caption{Electron bound to a charged superconducting ball.  }
%   \label{balls}
%\end{figure}

If we consider a similar example with charged spheres that are not so exotic as superconductors but just a ``strong'' dielectric as pyrolitic carbon with susceptibility $\chi\approx 40\times10^{-5}$ we can compute a new radius $R=3.6\mu$m, separation $r=12\mu$m and charge number $n=2476$.  Surprisingly, the large reduction is susceptibility does not eliminate the possibility of such states or create a large change in the scale of their size.  In the following we will see that small clusters have the most promise for stable collections.  

\section{Collections of Dimers}

It is now natural to ask what happens when we combine such pairs.  Unlike the case of real atoms where covalent bonds can exist with large barrier potentials to create new bonds, large collections of such dimers must give more generally coordinated collections with no clear distinction of which pair is bound.  The magnetite particles have the same sign charge so have a long range repulsion.  Similarly for the superconducting spheres.  A concern here is that the magnets can flip to be antiparallel and, at short ranges the magnetic attraction between them can overwhelm the electrostatic repulsion.  This reminds us that such an ensemble is, at best, metastable.  Indeed, one should be concerned that a pair of magnets could be drawn together with a superconducting sphere attenuating the net electrostatic repulsion leading to a kind of three-body collapse reminiscent of the losses in cold gas traps.  When the magnets and superconducting spheres are close, the form of the repulsion changes so that this may provide some brake on such a process.  

To obtain a binding likely to be stable enough in the case of many particles we need adequate repulsion to keep the magnetic spheres apart at close range so that the electrostatic repulsion is never overcome by oppositely oriented poles.  This distance must be on the order of the particle radii themselves so that multipolar fields can be set up in the superconductors strong enough to keep multiple such magnetic spheres from binding.  This can be estimated as above assuming $r\approx 2R$ and allowing $FE\gg Fg$ so that 
$R=9.4$nm and $n=6.6$.  Unfortunately, this does not give a large enough radius to be larger than the coherence length for any Type I superconductor.  In the case ceramic superconductors there are coherence lengths this small but the London penetration depth tends to be much larger.  It might be that some such material could be effective as a superconductor on such scales to give a strong repulsive force and make the plasma stable.  
Since it is unclear if such an Earnshaw plasma of clusters can exist stably for any particular material, we give a few properties that would characterize such a collection but do not develop it extensively.  
These pairs of ``atoms'' can have a broad range of internal excitations so there is no reasonable chance of any two being ``identical.''  The quantum pressure they would exert would be negligible in any case.  

We can compute the pressure of the plasma in our above example at $T=0$ by computing the weight of each sphere per area.  Approximating the particle separation as $2R$ we can assume that the density is one eighth that of the solid material and the pressure is determined by the energy density of the fields between them.  
%as unchanged from the singe pair case, $r$, we see that $P=6\times10^{-4}$N/m$^{2}$.  This tiny pressure belies the fact that, since the gas density is so low $\rho_{net}=3.2$kg/m$^{3}$, we can tune the interactions so that it can be stable for many layers deep.  Such an exotic gas might be an ideal probe of diffusion and gravitational scale height effects.  
The ``temperature'' of such a gas can be excited by jiggling from mechanical or field induced means.  Due to the poor coupling of the translational motions and the internal atomic motions such a damping will take a very long time.  This allows us to obtain very high temperatures with thermal velocities that are very small.  In our previous example $v_{th}\approx$1cm/s despite $T_{b}\sim10^{5}$K.  
For spherical bodies, the transfer of linear to angular momentum motion can be quite slow.  If we introduce angular bodies, this can be enhanced and lead to a specific heat of $c_{V}=\frac{5}{2}k_{B}T$.  There is a collective binding between the particles so we expect analogs to surface tension and vapor pressure.

\section{Conclusions}
It is rare to find thermodynamic systems that have dynamics that are slow enough to be tracked and observed.  Some may argue diffusion of large particles qualify but they are driven by small motions where this is not possible.  The existence of such classically bound pairs, and possibly plasma, strangely seems to be a novel realization given the advances that have taken place in the quantum world.  Even more interesting is that such pairs seem to be actually makable and fairly easily so.  

Potential uses of such pairs range from mesoscopic probes of quantum-classical transitions to granular-fluid transitions and as imageable slow thermodynamic systems that can be driven far from equilibrium.  %Of course, these are hopeful speculations and, to resist the trend in science of proclaiming every advance as of enormous relevance and utility, they may be only that.  
Hopefully, the arguments here will convince someone skilled to make a few and see what their real possibilities are.  
%Sound and diffusion speeds are very slow.  Insulation to the internal atomic degrees of freedom.  External field handles by polarizability.  Means to transport delicate small objects with insulated gaps from phonon pathways.  Granular-fluid transition uncomplicated by friction.  Geometric obstruction is all that exists.  Optical traps.  Example of hydro in exotic potentials.  
Should these be stable and resilient as suggested there are a number of ways they can be extended to richer objects.  By virtue of these being classical objects, we could consider modifying the spheres to become 
hinged and flexibly joined structures.  This allows a controlled introduction of internal degrees of freedom.  Interestingly, these bodies can all be viewed as distinguishable yet thermodynamics based on kinetics almost certainly exists and no Gibbs paradox seems relevant.  

%well defined surfaces

One might naturally wonder if we could utilize the diamagnetic response to bind an atom to a charged superconducting cluster in a kind of mesoscopic analog of the Penning trap.  It seems that the electrostatic force is too strong even for a single electron's worth of charge to make a stable system this way.  
In the quantum regime there are a large zoo of unusual bound states with diffuse electron clouds as in Rydberg matter and dipole bound anions.   Hyperfine interactions in dilute gases can be tuned magnetically to give Efimov states and a crossover transition between BCS and BEC behavior \cite{Dolfovo}.  Geonium \cite{Geonium} is a kind of simple atom where one ``half'' of it is the earth itself and the binding force is gravity.  Here we have a classical collection where the delocalization time is very long.  This does not mean that it has to be effectively infinite.  The quantum-classical transition is of great interest due to interest in quantum computing and the stability of such complicated entangled collections.  Such small particles in intense isolation could slowly delocalize.  One could also consider manufacturing them from a delocalized gas in a trap capable of holding both the gas and the transitional larger clusters forming the solid despite the impulses of photon emission and evaporation.  If the ejected particles were not allowed to collapse the resulting cluster should remain in a delocalized state.  Macroscopic superposition effects then could be measured and provide more fuel to the subject of quantum measurement.

\end{document}